\documentstyle[epsfig,proceedings]{crckapb} 

\newcommand{\beq}{\begin{equation}}
\newcommand{\eeq}{\end{equation}}
\newcommand{\bd}{\begin{displaymath}}
\newcommand{\ed}{\end{displaymath}}
 
\newcommand{\p}{\partial}        
\newcommand{\rmd}{{\rm d}}

\newcommand{\aaa}[1]{{\it Astron.\ Astrophys.} {\bf #1}}

\newcommand{\apj}[1]{{\it Astrophys. J.} {\bf #1}}

\newcommand{\ass}[1]{{\it Ap.\ Space Sci.} {\bf #1}}

\newcommand{\mnras}[1]{{\it Mon.\ Not.\ R.\ Astron.\ Soc.} {\bf #1}}   
\newcommand{\nature}[1]{{\it Nature} {\bf #1}}

\newcommand{\pasj}[1]{{\it Publ.\ Astr.\ Soc.\ Japan} {\bf #1}}        
\newcommand{\pasp}[1]{{\it Publ.\ Astr.\ Soc.\ Pac.} {\bf #1}}

\newcommand{\gapprox}{\;\rlap{\lower 2.5pt
             \hbox{$\sim$}}\raise 1.5pt\hbox{$>$}\;}
\newcommand{\lapprox}{\;\rlap{\lower 2.5pt
             \hbox{$\sim$}}\raise 1.5pt\hbox{$<$}\;}
\newcommand{\bfg}[1]{\setbox0=\hbox{#1}%
  \kern-.025em\copy0\kern-\wd0
  \kern.05em\copy0\kern-\wd0
  \kern-.025em\raise.0433em\box0}

\begin{opening}
  \title{Accretion Disks}  
  \author{H.C. Spruit}
  \institute{Max-Planck-Institut f\"ur Astrophysik\\ Postfach 1523, D-85740
Garching, Germany}
\end{opening}
\begin{document}
\begin{abstract}
In this lecture the basic theory of accretion disks is
introduced, with emphasis on aspects relevant for X-ray binaries and
Cataclysmic Variables.

To appear in `The neutron star black hole connection' (NATO ASI Elounda 1999,
eds. C. Kouveliotou and V. Connaughton).

 \keywords neutron stars, black holes, Caraclysmic Variables, accretion: accretion disks
\end{abstract}

\section{Introduction}
Accretion disks are inferred to exist in objects of very different scales: km to millions of km in low Mass X-ray Binaries (LMXB) and Cataclysmic Variables (CV), solar radius-to-AU scale in  protostellar disks, and AU-to-parsec scales for the disks in Active Galactic Nuclei (AGN). 

An interesting observational connection exists between accretion disks and 
jets (such as the spectacular jets from AGN and protostars), and outflows 
(the `CO-outflows' from protostars and possibly the `broad-line-regions' in 
AGN).  Lacking direct (i.e. spatially resolved) observations of disks, theory has
tried to provide models, with varying degrees of success. Uncertainty still exists with respect to some basic questions. In this situation, progress
made by observations or modeling of a particular class of objects is likely to
have direct impact for the understanding of other objects, including the enigmatic connection with jets.

In this lecture I concentrate on the more basic aspects of accretion disks, but an attempt is made to mention topics of current interest, such as magnetic viscosity, as well. Emphasis is on those aspects of  accretion disk theory that connect to the observations of LMXB and CV's. 
For other reviews on the basics of accretion disks, see Pringle (1981),
Treves et al. (1988).  For a more in-depth treatment, see the textbook by Frank et al. (1992).

\section{Accretion: general}

Gas falling into a point mass potential
\bd \Phi=-{GM\over r} \ed
from a distance $r_0$ to a distance $r$ converts gravitational into kinetic
energy, by an amount $\Delta\Phi=GM(1/r-1/r_0)$. For simplicity, assuming that the starting distance is large, $\Delta\Phi=GM/r$. If the gas is then brought to rest, for example at the surface of a star, the amount of energy $e$ dissipated per unit mass is  
\bd e={GM\over r} \qquad (\mathrm rest) \ed
or, if it goes into a circular Kelper orbit at distance $r$:
\bd e={1\over 2}{GM\over r} \qquad (\mathrm orbit). \ed
The dissipated energy goes into internal energy of the gas, and into radiation
which escapes to infinity (usually in the form of photons, but neutrino losses
can also play a role).

\subsection{Adiabatic accretion}
Consider first the case when radiation losses are neglected. This is 
{\it adiabatic} accretion. For an ideal gas with constant ratio of specific
heats $\gamma$, the internal energy per unit mass is
\bd e={P\over (\gamma -1)\rho}. \ed
With the equation of state
\beq P={\cal R}\rho T/\mu \label{es}\eeq
where ${\cal R}$ is the gas constant, $\mu$ the mean atomic weight per particle, we find the temperature of the gas after the dissipation has taken place (assuming that the gas goes into a circular orbit): 
\beq T={1\over 2}(\gamma-1)T_{\rm vir}, \label{ttv}\eeq
where $T_{\rm vir}$, the {\it virial temperature} is given by
\bd T_{\rm vir}={GM\mu\over{\cal R}r}. \ed
In an atmosphere with temperature near $T_{\rm vir}$, the sound speed is close
to the escape speed from the system, the hydrostatic pressure scale height is
of the order of $r$, and such an atmosphere may evaporate on a relatively short
time scale in the form of a stellar wind.

A simple example is {\it spherical} adiabatic accretion (Bondi, 1952). An important result is that such accretion is possible only
if $\gamma\le 5/3$. The larger $\gamma$, the larger the
temperature in the accreted gas (eq.~\ref{ttv}), and beyond a critical value
the temperature is too high for the gas to stay bound in the potential. A
classical situation where adiabatic and roughly spherical accretion takes
place is a supernova implosion: when the central temperature becomes high
enough for the radiation field to start desintegrating nuclei, $\gamma$ drops
and the envelope collapses onto the forming neutron star via a nearly static
accretion shock. Another case are Thorne-Zytkow objects (e.g. Cannon et al. 1992), where $\gamma$ can drop to low values due to pair creation,
initiating an adiabatic accretion onto the black hole.

Adiabatic spherical accretion is fast, taking place on the dynamical or free
fall time scale 
\beq \tau_{\rm d}=r/v_{\rm K}=(r^3/GM)^{1/2}, \eeq
where $v_{\rm K}$ is the Kepler orbital velocity.

When radiative loss becomes important, the accreting gas can stay cool
irrespective of the value of $\gamma$, and Bondi's critical value $\gamma=5/3$
plays no role. With such losses, the temperatures of accretion disks are usually
much lower than the virial temperature.  The optical depth of the accreting
flow increases with the accretion rate $\dot M$. When the optical depth
becomes large enough so that the photons are `trapped' in the flow, the
accretion just carries them in, together with the gas (Rees 1978, Begelman
1979).  Above a certain critical rate $\dot M_{\rm c}$,  accretion is therefore adiabatic.

\subsection{The Eddington Limit}
Objects of high luminosity have a tendency to blow their atmospheres away due
to the radiative force exerted when the outward traveling photons are scattered
or absorbed. Consider a volume of gas on which a flux of photons is incident
from one side. Per gram of matter, the gas presents a scattering (or
absorbing) surface area of $\kappa$ cm$^2$. The force exerted by the
radiative flux $F$ on one gram is $F\kappa/c$. The force of gravity pulling
back on this one gram of mass is $GM/r^2$. The critical flux at which the two
forces balance is
\beq F_{\rm E}={c\over\kappa}{GM\over r^2} \label{fe}\eeq
Assuming that the flux is spherically symmetric, this can be converted into a
critical luminosity
\beq L_{\rm E}=4\pi GMc/\kappa ,\label{le}\eeq
the Eddington luminosity (e.g. Rybicki and Lightman, 1979). If the gas is fully
ionized, its opacity is dominated by electron scattering, and for solar
composition $\kappa$ is then of the order $0.3$ cm$^2$/g (about a factor 2
lower for fully ionized helium, a factor up to $10^3$ higher for partially
ionized gases). With these assumptions, 
\bd L_{\rm E}\approx 1.7\,10^{38}{M\over M_\odot}~~{\mathrm erg/s}\approx 4\,10^4{M\over
M_\odot}~~L_\odot\ed
If this luminosity results from accretion, it corresponds to the Eddington
accretion rate $\dot M_{\rm E}$:
\beq {GM\over r}\dot M_{\rm E}=L_{\rm E}\quad\rightarrow\quad\dot M_{\rm E}=4\pi rc/\kappa
\label{me}.\eeq
Whereas $L_{\rm E}$ is a true limit that can not be exceeded by a static radiating
object except by geometrical factors of order unity (see chapter 10 in Frank
et al, 1992), no maximum exists on the accretion rate. For $\dot M>\dot M_{\rm E}$
the plasma is just swallowed whole, including the radiation energy in it (cf.
discussion in the preceding section). With $\kappa=0.3$:
\bd \dot M_{\rm E}\approx 1.3\,10^{18}r_6~~{\mathrm g/s}\approx 2\,10^{-8}
r_6~~M_\odot{\mathrm yr}^{-1}, \ed
where $r_6$ is the radius of the accreting object in units of 10 km.

\section{Accretion with Angular Momentum}
When the accreting gas has a zonzero angular momentum with respect to the
accreting object, it can not accrete directly. A new time scale appears,
the time scale for outward transport of angular momentum. Since this is in
general much longer than the dynamical time scale, much of what was said about spherical accretion needs modification for accretion with angular momentum.

Consider the accretion in a close binary consisting of a compact (white dwarf,
neutron star or black hole) primary of mass $M_1$ and a main sequence
companion of mass $M_2$. The mass ratio is defined as $q=M_2/M_1$ (note:
$q$ is just as often defined the other way around).

\begin{figure}[htbp]
\mbox{}\hfill\epsfysize6cm\epsfbox{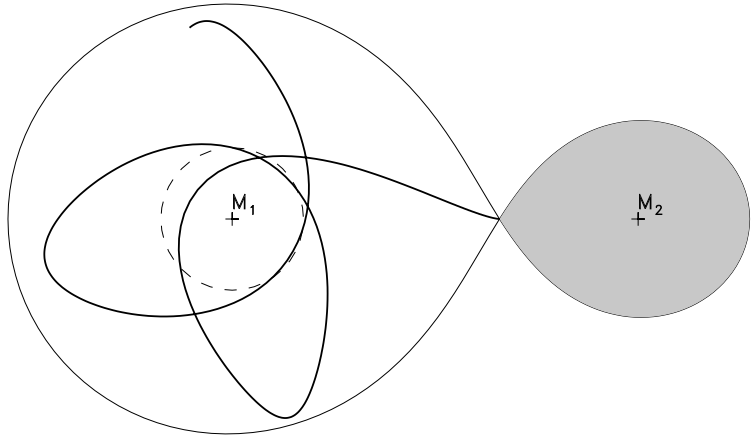}\hfill\mbox{}
 \caption[]{\label{rof}
Roche geometry for $q=0.2$, with free particle orbit from $L_1$ (as seen in a frame corotating with the orbit). Dashed: circularization radius.}   
\end{figure}

If  $M_1$ and $M_2$ orbit each other in a circular orbit and their separation
is $a$, the orbital frequency $\Omega$ is
\bd \Omega^2=G(M_1+M_2)/a^3 .\ed
The accretion process is most easily described in a coordinate frame that
corotates with this orbit, and with its origin in the center of mass. Matter
that is stationary in this frame experiences an effective potential, the
{\it Roche potential} (Ch. 4 in Frank, King and Raine, 1992), given by
\beq \phi_{\rm R}({\mathbf r})=-{GM\over r_1}-{GM\over r_2}-{1\over 2}\Omega^2r^2
\label{ro}\eeq 
where $r_{1,2}$ are the distances of point $\mathbf r$ to stars 1,2. Matter that
does {\it not} corotate experiences a very different force (due to the Coriolis
force). The Roche potential is therefore useful only in a rather limited
sense. For non-\-corotating gas intuition based on the Roche geometry is usually
confusing. Keeping in mind this limitation, consider the equipotential
surfaces of (\ref{ro}). The surfaces of stars $M_{1,2}$, assumed to corotate
with the orbit, are equipotential surfaces of (\ref{ro}). Near the centers of
mass (at low values of $\phi_{\rm R}$) they are unaffected by the other star, 
at higher $\Phi$ they are distorted and at a critical value $\Phi_1$ the two
parts of the surface touch. This is the critical Roche surface $S_1$ whose two
parts are called the Roche lobes. Binaries lose angular momentum through
gravitational radiation and a magnetic wind from the secondary (if it has
a convective envelope). Through this loss the separation between the
components decreases and both Roche lobes decrease in size. Mass transfer
starts when $M_2$ fills its Roche lobe, and continues as long as the angular
momentum loss from the system lasts. A stream of gas then flows through the point of
contact of the two parts of $S_1$, the inner Lagrange point $L_1$. If the
force acting on it were derivable entirely from (\ref{ro}) the gas would just
fall in radially onto $M_1$. As soon as it moves however, it does not corotate
any more and its orbit under the influence of the Coriolis force is different
(Fig.~\ref{rof}).

Since the gas at $L_1$ is very cold compared with the virial temperature,
its sound speed is small compared with the velocity it gets after only a
small distance from $L_1$. The flow into the Roche lobe of $M_1$ is
therefore highly {\it supersonic}. Such hypersonic flow is essentially
ballistic, that is, the stream flows along the path taken by free particles. 

Though the gas stream on the whole follows an orbit close to that of a free
particle, a strong shock develops at the point where the orbit intersects
itself. [In practice shocks already develop shortly after passing the
pericenter at $M_1$, when the gas is decelerated again. Supersonic flows that
are decelerated by whatever means in general develop shocks (e.g. Courant and
Friedrichs 1948, Massey, 1968). The effect can be seen in action in the movie
published in R\'o\.zyczka and Spruit, 1993]. After this, the gas settles into
a ring, into which the stream continues to feed mass. If the mass ratio $q$ is
not too small this ring forms fairly close to $M_1$. An approximate value for
its radius is found by noting that near $M_1$ the tidal force due to the
secondary is small, so that the angular momentum of the gas with respect to
$M_1$ is approximately conserved. If the gas continues to conserve angular
momentum while dissipating energy, it settles into the minimum energy orbit
with the specific angular momentum $j$ of the incoming stream. The radius of
this orbit, the {\it circularization radius} $r_{\rm c}$ is determined from 
\bd (GM_1r_{\rm c})^{1/2}=j. \ed 
The value of $j$ is found by a simple integration of the orbit starting at
$L_1$ and measuring $j$ at some point near pericenter. In units of the
orbital separation $a$, $r_{\rm c}$ and the distance $r_{{\rm L}1}$ from $M_1$ to $L_1$
are functions of the mass ratio only. As an example for $q=0.2$,
$r_{{\rm L}1}\approx 0.66a$ and the circularization radius $r_{\rm c}\approx 0.16a$. In practice the ring forms somewhat outside $r_{\rm c}$, because there is some angular momentum redistribution in the shocks that form at the impact of the stream on the ring.

The evolution of the ring depends critically on nature and strength of the
angular momentum transport processes. If sufficient `viscosity' is present it
spreads inward and outward to form a disk.

At the point of impact of the stream on the disk the energy dissipated is
a significant fraction of the orbital kinetic energy, hence the gas heats up to
a significant fraction of the virial temperature. For a typical system with
$M_1=1 M_\odot$, $M_2=0.2 M_\odot$ having an orbital period of 2 hrs, the
observed size of the disk (e.g. Wood et al. 1989b, Rutten et al. 1992)
$r_{\rm d}/a\approx 0.3$, the orbital velocity at $r_{\rm d}$ about 900 km/s, the virial
temperature at $r_{\rm d}$ is $10^8$K. The actual temperatures at the impact point
are much lower, due to rapid cooling of the shocked gas. Nevertheless the
impact gives rise to a prominent `hot spot' in many systems, and an overall
heating of the outermost part of the disk. 

\section{Thin disks: properties}
\subsection{Flow in a cool disk is supersonic}
Ignoring viscosity, the equation of motion in the potential of a point mass is
\beq {\p{\mathbf v}\over\p t}+{\mathbf v}\cdot\nabla{\mathbf v}=-{1\over\rho}\nabla P-
{GM\over r^2}{\mathbf{\hat r}}, \eeq
where $\mathbf{\hat r}$ is a unit vector in the spherical radial direction $r$.
To compare the order of magnitude of the terms, choose a position $r_0$ in the
disk, and choose as typical time and velocity scales the orbital time scale
$\Omega_0^{-1}=(r_0^3/GM)^{1/2}$ and velocity $\Omega_0r_0$. The pressure gradient term is 
\bd {1\over\rho}\nabla P={{\cal R}\over\mu}T\nabla\ln P. \ed
In terms of the dimensionless quantities 
\bd \tilde r=r/r_0, \qquad \tilde v=v/(\Omega_0r_0), \ed 
\bd \tilde t=\Omega_0t, \qquad \tilde\nabla=r_0\nabla, \ed
the equation of motion is then
\beq {\p{\mathbf {\tilde v}}\over\p \tilde t}+{\mathbf {\tilde v}}\cdot \tilde\nabla 
{\mathbf{\tilde v}}=-{T\over T_{\rm vir}}\tilde\nabla\ln P-{1\over\tilde r^2}
{\mathbf{\hat r}}.
\eeq
All terms and quantities in this equation are of order unity by the
assumptions made, except the pressure gradient term which has the coefficient
$T/T_{\rm vir}$. If cooling is important, so that $T/ T_{\rm vir}\ll 1$, the pressure
term is negligible to first approximation, and vice versa. Equivalent
statements are also that the gas moves hypersonically on nearly Keplerian
orbits, and that the disk is thin, as is shown next.

\subsection{Disk thickness}
The thickness of the disk is found by considering its equilibrium in the
direction perpendicular to the disk plane. In an axisymmetric disk, using
cylindrical coordinates ($\varpi,\phi,z$), measure the forces at a point ${\mathbf r}_0$ ($\varpi,\phi,0$) in the midplane, in a frame rotating with the Kepler rate $\Omega_0$ at that point. The gravitational acceleration
$-GM/r^2\,{\mathbf{\hat r}}$ balances the centrifugal acceleration $\Omega_0^2$\bfg{$\varpi$}
at this point, but not at some distance $z$ above it because gravity and
centrifugal acceleration work in different directions. Expanding both
accelerations near ${\mathbf r}_0$, one finds a residual acceleration toward the midplane of magnitude \bd g_z=-\Omega_0^2z. \ed
Assuming again an isothermal gas, the condition for equilibrium in the $z$
direction under this acceleration yields a hydrostatic density distribution
\bd \rho=\rho_0(\varpi)\exp\left(-{z^2\over 2H^2}\right). \ed
$H$, the {\it scale height} of the disk, is given in terms of the isothermal
sound speed $c_{\rm s}=({\cal R}T/\mu)^{1/2}$ by
\bd H=c_{\rm s}/\Omega_0. \label{h}\ed
We define $\delta\equiv H/r$, the {\it aspect ratio} of the disk, and find that
it can be expressed in several equivalent ways: 
\bd
\delta={H\over r}={c_{\rm s}\over \Omega r}=M^{-1}=\left ({T\over T_{\rm vir}}
\right) ^{1/2}, \ed
where $M$ is the Mach number of the orbital motion. 

\subsection{Viscous spreading}
The shear flow between neighboring Kepler orbits in the disk causes friction
due to viscosity. The frictional torque is equivalent to exchange of angular
momentum between these orbits. But since the orbits are close to Keplerian, a
change in angular momentum of a ring of gas also means it must change its disctance from the central mass. If the angular momentum is increased, the ring moves to a larger
radius. In a thin disk angular momentum transport (more precisely a
nonzero divergence of the angular momentum flux) therefore automatically
implies redistribution of mass in the disk. 

A simple example (L\"ust 1952, see also Lynden-Bell and Pringle 1974) is a
narrow ring of gas at some distance $r_0$. If at $t=0$ this ring is released to
evolve under the viscous torques, one finds that it first spreads into an
asymmetric hump with a long tail to large distances. As $t\rightarrow\infty$ the hump flattens in such a way that almost all the {\it mass}
of the ring is accreted onto the center, while a vanishingly small fraction of
the gas carries almost all the {\it angular momentum} to infinity. As a result of this
asymmetric behavior essentially all the mass of a disk can accrete, even if
there is no external torque to remove the angular momentum.

\subsection{Observations of disk viscosity}
Evidence for the strength of the angular momentum transport processes in
disks comes from observations of variability time scales. This evidence is
not good enough to determine whether the processes really have the same
effect as a viscosity, but if this is assumed, estimates can be made
of the magnitude of the viscosity.

Cataclysmic Variables give the most detailed information. These are binaries
with white dwarf (WD) primaries and (usually) main sequence companions (for
reviews see Meyer-Hofmeister and Ritter 1993, Cordova 1995, Warner 1995). A subclass of
these systems, the Dwarf Novae, show semiregular outbursts. In the currently
most developed theory, these outbursts are due to an instability in the disk
(Smak 1971, Meyer and Meyer-Hofmeister 1981, for recent references see, King 1995, Hameury et al. 1998). The outbursts are episodes of enhanced
mass transfer of the disk onto the primary, involving a significant part of the
whole disk. The decay time of the burst is thus a measure of the viscous time
scale of the disk (the quantitative details depend on the model, see Cannizzo
et al. 1988, Hameury et al. 1998): 
\bd t_{\rm visc}=r_{\rm d}^2/\nu ,\ed
where $r_{\rm d}$ is the size of the disk. With decay times on the order of days,
this yields viscosities of the order $10^{15}$ cm$^2$/s, about 14 orders of
magnitude above the microscopic viscosity of the gas.

Other evidence comes from the inferred time scale on which disks around
protostars disappear, which is of the order of $10^7$ years (Strom et al,
1993).

\subsection{$\alpha$-parametrization}
The process responsible for such a large viscosity has not been identified
with certainty yet. Many processes have been proposed, some of which demonstrably work,
though often not with an efficiency as high as the observations of CV
outbursts seem to indicate. Other ideas, such as certain turbulence models, do not have much predictive power and are based on ad-hoc assumptions about hydrodynamic instabilities in disks. In order to
compare the viscosities in disks under different conditions, one introduces a
dimensionless vsicosity $\alpha$:
\beq \nu=\alpha{c_{\rm s}^2\over\Omega}, \label{alf}\eeq
where $c_{\rm s}$ is the isothermal sound speed as before. The quantity $\alpha$
was introduced by Shakura and Sunyaev (1973), as a way of parametrizing our
ignorance of the angular momentum transport process (their definition is based
on a different formula however, and differs by a constant of order unity).

How large can the value of $\alpha$ be, on theoretical grounds? As a simple
model, let's assume that the shear flow between Kepler orbits is unstable to
the same kind of shear instabilities found for flows in tubes, channels, near
walls and in jets. These instabilities occur so ubiquitously that the fluid
mechanics community considers them a natural and automatic consequence (e.g.
DiPrima and Swinney 1981, p144 2nd paragraph) of a high Reynolds number:
\bd {\mathrm Re}={LV\over\nu} \ed
where $L$ and $V$ are characteristic length and velocity scales of the flow.
If this number exceeds about 1000 (for some forms of instability much less),
instability and turbulence are generally observed. It has been argued
(e.g. Zel'dovich 1981) that for this reason hydrodynamic turbulence is the cause of disk viscosity. Let's look at the consequences of this assumption. If an eddy of radial length scale $l$ develops due to shear instability, it will rotate at a rate given by the rate of shear, $\sigma$, in the flow, here 
\bd \sigma=r{\p\Omega\over\p r}\approx-{3\over 2}\Omega. \ed
The velocity amplitude of the eddy is $V=\sigma l$, and a field of such eddies
produces a turbulent viscosity of the order (leaving out numerical factors of
order unity)
\beq \nu_{\rm turb}=l^2\Omega  .\eeq
In compressible flows, there is a maximum to the size of the eddy set by
causality considerations. The force that allows an instability to form an
overturning eddy is the pressure, which transports information about the flow
at the sound speed. The eddies formed by a shear instability can therefore
not move faster than $c_{\rm s}$, hence their size does not exceed
$c_{\rm s}/\sigma\approx H$. At the same time, the largest eddies formed also have
the largest contribution to the turbulent viscosity. Thus we should expect
that the turbulent viscosity is given by eddies with size of the order $H$:
\bd \nu\sim H^2\Omega, \ed
or 
\bd \alpha\sim 1.\ed
Does hydrodynamical turbulence along these lines exist in disks? Unfortunately,
this question is still open, but current opinion is leaning toward the view that the angular momentum transport in sufficiently ionized disks is due a a small scale magnetic field (Shakura and Sunyaev 1973). This is discussed briefly in
section 8.

\section{Thin Disks: equations}
Consider a thin (= cool, nearly Keplerian, cf. section 4.2) disk, axisymmetric
but not stationary. Using cylindrical coordinates ($r,\phi,z$), (note that we
have changed notation from $\varpi$ to $r$ compared with section 4.2) we define
the {\it surface density} $\Sigma$ of the disk as  
\beq \Sigma=\int_{-\infty}^\infty\rho{\mathrm d}z\approx
2H_0\rho_0, \eeq  
where $\rho_0$, $H_0$ are the density and scaleheight at the midplane. The
approximate sign is used to indicate that the coefficient in front of $H$ in
the last expression actually depends on details of the vertical structure of
the disk. Conservation of mass, in terms of $\Sigma$ is given by 
\beq{\p\over\p t}(r\Sigma)+{\p\over\p r}(r\Sigma v_r)=0. \label{co}\eeq
(derived by integrating the continuity equation over $z$). Since the disk is
axisymmetric and nearly Keplerian, the radial equation of motion reduces to
\beq v_\phi^2={GM/ r}. \eeq
The $\phi$-equation of motion is
\beq
{\p v_\phi\over\p t}+v_r{\p v_\phi\over\p r}+{v_rv_\phi\over r}=F_\phi,\label{vphi}
\eeq
where $F_\phi$ is the azimuthal component of the viscous force.
By integrating this over height $z$ and using (\ref{co}),
one gets an equation for the angular momentum balance:
\beq 
{\p\over\p t}(r\Sigma\Omega r^2)+{\p\over\p r}(r\Sigma v_r\Omega r^2)=
{\p\over\p r}(Sr^3{\p\Omega\over\p r}), \label{ang}
\eeq
where $\Omega=v_\phi/r$, and 
\beq
S=\int_{-\infty}^\infty \rho\nu {\mathrm d}z\approx\Sigma\nu. \label{s}
\eeq
The second approximate equality in~(\ref{s}) holds if $\nu$ can be considered
independent of $z$.  The right hand side of (\ref{ang}) is the divergence of the 
viscous angular momentum flux, and is derived most easily with a physical 
argument, as described in, e.g. Pringle (1981) or Frank et al. (1992)\footnote{If
you prefer a more formal derivation, the fastest way is to consult Landau and
Lifshitz (1959) chapter 15 (hereafter LL). Noting that the divergence of the
flow vanishes for a thin axisymmetric disk, the viscous stress $\sigma$
becomes (LL eq. 15.3)  
\bd 
\sigma_{ik}=\eta\left({\p v_i\over\p x_k}+{\p v_k\over\p x_i} \right), 
\ed 
where $\eta=\rho\nu$. This can be written in cylindrical or spherical coordinates 
using LL eqs. (15.15-15.18). The viscous force is 
\bd 
F_i={\p\sigma_{ik}\over\p x_k}={1\over\eta}{\p\eta\over\p x_k}\sigma_{ik} +\eta\nabla^2 v_i, 
\ed
Writing the Laplacian in cylindrical coordinates, the viscous torque is then computed  from the $\phi$-component of the viscous force by multiplying by $r$, and is then integrated over $z$.}. 

Assume now that $\nu$ can be taken constant with height. For an isothermal disk ($T$ independent of $z$), this is equivalent to taking the viscosity parameter 
$\alpha$ independent of $z$. As long as we are not sure what causes the 
viscosity this is a reasonable simplification. Note, however, that recent numerical simulations of magnetic turbulence suggest that the effective 
$\alpha$, and the rate of viscous dissipation per unit mass, are higher near the 
disk surface than near the midplane. See the discussion in section 
\ref{source}. While eq (\ref{ang}) is still valid for rotation rates $\Omega$ deviating from Keplerian (only the integration over disk thickness must be justifiable), we now use the fact that $\Omega\sim r^{-3/2}$. Then
Eqs.~(\ref{co}-\ref{ang}) can then be combined into a single equation for $\Sigma$:
\beq  r{\p\Sigma\over\p t}=3{\p\over\p r}[r^{1/2}{\p\over\p r}(\nu
\Sigma r^{1/2})] .\label{dif} 
\eeq
Under the same assumptions, eq. (\ref{vphi}) yields the mass flux $\dot M$ at any point in the disk:
\beq \dot M=-2\pi r\Sigma v_r=6\pi r^{1/2}{\p\over\p r}(\nu\Sigma
r^{1/2}).\label{dm}\eeq 

Eq. (\ref{dif}) is the standard form of the {\it thin disk diffusion equation}. An
important conclusion from this equation is: in the thin disk limit, all the
physics which determines the time dependent behavior of the disk enters 
through one quantitity only, the viscosity $\nu$. This is the main attraction
of the thin disk approximation.

\subsection{Steady Thin Disks}
In a steady disk ($\p/\p t=0$) the mass flux $\dot M$ is constant through the
disk and equal to the accretion rate onto the central object. From (\ref{dm})
we get the surface density distribution:
\beq 
\nu\Sigma={1\over 3\pi}\dot M\left[1-\beta\left({r_i\over r}\right)^{1/2}
\right], \label{ns}
\eeq
where $r_i$ is the inner radius of the disk and $\beta$ is a parameter
appearing through the integration constant. It is related to the flux of
angular momentum $F_J$ through the disk: 
\beq F_J=-\dot M\beta\Omega_ir_i^2 \label{fj}, \eeq 
where $\Omega_i$ is the Kepler rotation rate at the inner edge of the disk.
If the disk accretes onto an object with a rotation rate $\Omega_*$ {\it less}
than $\Omega_i$, one finds (Shakura and Sunyaev, 1973, Lynden-Bell and
Pringle, 1974) that $\beta=1$, independent of $\Omega_*$. Thus the angular momentum flux (torque on the accreting star) is inward (spin-up) and equal to the accretion rate times the specific angular momentum at the inner edge of the disk. For stars rotating near their maximum rate
($\Omega_*\approx\Omega_i$) and for accretion onto magnetospheres, which can rotate faster than the disk, the situation is different (Sunyaev and Shakura 1977, Popham and Narayan 1991, Paczy\'nski 1991, Bisnovatyi-Kogan 1993). 

Accreting magnetospheres, for example, can {\it spin down} by interaction with the disk. This case has a surface density distribution (\ref{ns}) with $\beta<1$ (see also Spruit and Taam 1993). The angular momentum flux is then outward, and the accreting star spins down. This is possible even when the interaction between the disk and the magetosphere takes place {\it only} at the inner edge of the disk. Magnetic torques due interaction with the magetosphere
may exist at larger distances in the disk as well, but are not necessary for creating an outward angular momentum flux. Recent numerical simulations of disk-magnetosphere interaction (Miller and Stone 1997) give an interesting new view of how such interaction may take place, and suggests it happens very differently from what is assumed in previous `standard' models.

\subsection{Disk Temperature}
In this section I assume accretion onto not-too-rapidly rotating objects, so
that $\beta=1$. The surface temperature of the disk, which determines how
much energy it loses by radiation, is governed primarily by the energy
dissipation rate in the disk, which in turn is given by the accretion rate.
From the first law of thermodynamics we have
\beq \rho T{{\mathrm d}S\over {\mathrm d}t}=-{\mathrm div}{\mathbf F}+Q_{\rm v}, \eeq
where $S$ the entropy per unit mass, $\mathbf F$ the heat flux (including
radiation and any form of `turbulent' heat transport), and $Q_{\rm v}$ the viscous
dissipation rate. For changes which happen on time scales longer than the
dynamical time $\Omega^{-1}$, the left hand side is small compared with the
terms on the right hand side. Integrating over $z$, the divergence turns
into a surface term and we get 
\beq 
2\sigma_{\rm r}T_{\rm s}^4=\int_{-\infty}^\infty Q_{\rm v}{\mathrm d}z,\label{bal}
\eeq 
where $T_{\rm s}$ is the surface temperature of the disk, $\sigma_{\rm r}$ is Stefan-Boltzmann's radiation constant $\sigma_{\rm r}=a_{\rm r}c/4$, and the factor 2 comes about because the disk has 2  radiating surfaces (assumed to radiate like black bodies).
Thus the energy balance is {\it local} (for such slow changes): what is
generated by viscous dissipation inside the disk at any radius $r$ is also
radiated away from the surface at that position. The viscous dissipation rate
is equal to $Q_{\rm v}=\sigma_{ij}\p v_i/\p x_j$, where $\sigma_{ij}$
is the viscous stress tensor (see footnote in section 5), and this works out\footnote{using, e.g. LL eq. 16.3}
to be 
\beq Q_{\rm v}=9/4\,\Omega^2\nu\rho.\label{qv}\eeq 
Eq.~(\ref{bal}), using (\ref{ns}) then  gives the surface temperature in terms of the accretion rate: 
\beq 
\sigma_{\rm r} T_{\rm s}^4={9\over 8}\Omega^2\nu\Sigma={GM\over r^3}{3\dot
M\over 8\pi} \left[1-\left({r_i\over r}\right)^{1/2}\right]. \label{st}
\eeq
This shows that the surface temperature of the disk, at a given distance $r$
from a steady accreter, depends {\it only} on the product $M\dot M$, and not on
the highly uncertain value of the viscosity. For $r\gg r_i$ we have
\beq T_{\rm s}\sim r^{-3/4}. \eeq 

These considerations only tells us about the surface temperature. The
internal temperature in the disk is quite different, and depends on the
mechanism transporting energy to the surface. Because it is the internal
temperature that determines the disk thickness $H$ (and probably also the
viscosity), this transport needs to be considered in some detail for
realistic disk models. This involves a calculation of the vertical structure of
the disk. Because of the local (in $r$) nature of the balance between
dissipation and energy loss, such calculations can be done as a grid of models
in $r$, without having to deal with exchange of energy between neighboring
models. Schemes borrowed from stellar structure computations are used (e.g.
Meyer and Meyer-Hofmeister 1982, Pringle et al. 1986, Cannizzo et al.
1988). 

An approximation to the temperature in the disk can be found  when a number
of additional assumptions is made.  As in stellar interiors, the energy transport is radiative rather than convective at high temperatures. Assuming local thermodynamic equilibrium (LTE, e.g. Rybicki and Lightman 1979), the temperature structure of a radiative atmosphere is given, in the Eddington approximation by:
\beq {\rmd\over\rmd\tau} \sigma_{\rm r} T^4={3\over 4}F.\label{edd}\eeq
The boundary condition that there is no incident flux from outside the atmosphere yields the approximate condition
\beq \sigma_{\rm r} T^4(\tau=2/3)=F, \label{sb}\eeq
where $\tau=\int_z^\infty\kappa\rho \rmd z$ is the optical depth at geometrical depth $z$, and $F$ the energy flux through the atmosphere. Assuming that most
of heat is generated near the midplane (which is the case if $\nu$ is
constant with height), $F$ is approximately constant with height and equal to
$\sigma_{\rm r}T^4_{\rm s}$, given by (\ref{st}). Eq (\ref{edd}) then yields
\beq \sigma_{\rm r} T^4={3\over 4}(\tau+{2\over 3})F.\label{t4}\eeq
Approximating the opacity $\kappa$ as constant with $z$, the optical depth at the midplane is $\tau=\kappa\Sigma/2$. If $\tau\gg1$, the temperature at the midplane is then: 
\beq 
T^4={27\over 64}\sigma_{\rm r}^{-1}\Omega^2\nu\Sigma^2\kappa .\label{ts}
\eeq 
With the equation of state (\ref{es}), valid when radiation pressure is
small, we find for the disk thickness, using (\ref{ns}):  
\begin{eqnarray} 
{H\over r}=&({\cal R}/\mu)^{2/5}\left({3\over 64\pi^2\sigma_{\rm r}} \right)^{1/10}(\kappa/\alpha)^{1/10}(GM)^{-7/20}r^{1/20}(f\dot M)^{1/5} \cr
 =&5\,10^{-3}\alpha^{-1/10} r_6^{1/20} \left({M/M_\odot}\right)^{-7/20}
(f\dot M_{16})^{1/5}, \qquad (P_r\ll P)\label{hr} 
\end{eqnarray}
where $r_6=r/(10^6$ cm), $\dot M_{16}=\dot M/(10^{16}$g/s), and
\bd f=1-\left({r_i/ r}\right)^{1/2}. \ed
From this we conclude that: i) the disk is thin in X-ray binaries, $H/r<0.01$, ii) the disk thickness is relatively insensitive to the parameters,
especially $\alpha$, $\kappa$ and $r$. It must be stressed, however, that this
depends fairly strongly on the assumption that the energy is dissipated in the
disk interior. If the dissipation takes place close to the surface [such as in
some magnetic reconnection models (Haardt et al. 1994, Di Matteo et al. 1999 and references therein)], the internal disk temperature will be much closer to the surface temperature. The midplane temperature and $H$ are even smaller in such disks than calculated from (\ref{hr}).

The viscous dissipation rate per unit area of the disk, $W_{\rm v}=(9/4)\Omega^2\nu\Sigma$ [cf. eq. \ref{st})] can be compared with the local rate $W_{\rm G}$ at which gravitational energy is liberated in the accretion flow. Since half the gravitational energy stays in the flow as orbital motion, we have
\beq W_{\rm G}= {1\over 2\pi r}{GM\dot M\over 2 r^2},\eeq
so that
\beq W_{\rm v}/W_{\rm G}=3f=3[1-(r_{\rm i}/r)^{1/2}]. \eeq
At large distances from the inner edge, the dissipation rate is {\em 3 times larger than the rate of gravitational energy release}. This may seem odd, but becomes understandable when it is realized that there is a significant flux of energy through the disk associated with the viscous stress\footnote{See LL section 16}. 
Integrating the viscous energy dissipation over the whole disk, one finds 
\beq \int_{r_i}^\infty 2\pi r W_{\rm v} \rmd r={GM\dot M\over 2 r_{\rm i}},\eeq 
as expected. That is, globally, but not locally, half of the gravitational energy is radiated from the disk while the other half remains in the orbital kinetic energy of the accreted material. 

What happens to this remaining orbital energy depends on the nature of the accreting object. If the object is a slowly rotating black hole, the orbital energy is just swallowed by the hole. If it has a solid surface, the orbiting gas slows down until it corotates with the surface, dissipating the orbital energy into heat in a boundary layer. Unless the surface rotates close to the orbital rate (`breakup'), the energy released in this way is of the same order as the total energy released in the accretion disk. The properties of this boundary layer are therefore crucial for accretion onto neutron stars and white dwarfs. See also section \ref{bl} and  Inogamov and Sunyaev (1999, and elsewhere in this volume).

\subsection{Radiation pressure dominated disks}
In the inner regions of disks in XRB, the radiation pressure can dominate over
the gas pressure, which results in a different expression for the disk
thickness. The total pressure $P$ is 
\beq P=P_r+P_g={1\over 3}aT^4+P_g.\eeq
Defining a `total sound speed' by $c_{\rm t}^2=P/\rho$ the relation
$c_{\rm t}=\Omega H$ still holds. For $P_r\gg P_g$ we get from (\ref{ts}), with
(\ref{st}) and $\tau\gg 1$:
\bd c H={3\over 8\pi}\kappa f\dot M, \ed
(where the rather approximate relation $\Sigma=2H\rho_0$ has been used). Thus,
\beq 
{H\over R}\approx{3\over 8\pi}{\kappa\over cR}f\dot M ={3\over 2}f{\dot M\over\dot M_{\rm E}} , \label{he} 
\eeq
where $R$ is the stellar radius and $\dot M_{\rm E}$ the Eddington rate for this
radius. It follows that the disk becomes thick near the star, if the accretion
rate is near Eddington (though this is mitigated somewhat by the decrease of
the factor $f$). Accretion near the Eddington limit is evidently not geometrically thin any more. In addition, other processes such as angular momentum loss by `photon drag' have to be taken into account. 

\subsection{Time scales in a disk}
Three locally defined time scales play a role in thin disks. The dynamical time scale $t_{\rm d}$ is the orbital time scale:
\beq t_{\rm d}=\Omega^{-1}=({GM/ r^3})^{-1/2}. \eeq
The time scale for radial drift through the disk over a distance of order $r$ is the viscous time scale:
\beq 
t_{\rm v}=r/(-v_r)={2\over 3}{rf\over\nu}={2f\over 3\alpha\Omega}({r\over H})^2,
\eeq
(using (\ref{dm} and (\ref{ns}), valid for steady accretion). Finally, there are {\em thermal} time scales. If $E_{\rm t}$ is the thermal energy content (enthalpy) of the disk per unit of surface area, and $W_{\rm v}= (9/4)\Omega^2 \nu\Sigma$ the heating rate by viscous dissipation, we can define a heating time scale:
\beq t_{\rm h}=E_{\rm t}/W_{\rm v}. \eeq
In the same way, a cooling time scale is defined by the energy content and the radiative loss rate:
\beq t_{\rm c}=E_{\rm t}/(2\sigma_{\rm r} T_{\rm s}^4).\eeq
For a thin disk, the two are equal since the viscous energy dissipation is locally balanced by radiation from the two disk surfaces. [In thick disks (ADAFs), this balance does not hold, since the advection of heat with the accretion flow is not negligible. In ADAFs, $t_{\rm c}> t_{\rm h}$ (see elsewhere in this volume)].
Thus, we can replace both time scales by a single thermal time scale $t_{\rm t}$, and find, with (\ref{qv}):
\beq 
t_{\rm t}={1\over W_{\rm v}}\int_{-\infty}^\infty{\gamma P\over\gamma -1}\rmd z,
\eeq
where the enthalpy of an ideal gas of constant ratio of specific heats $\gamma$ has been used. Leaving out numerical factors of order unity, this yields
\beq t_{\rm t}\approx{1\over\alpha\Omega}.\eeq
That is, the thermal time scale of the disk is independent of most of the disk properties and of the order $1/\alpha$ times longer than the dynamical time scale. This independence is a consequence of the $\alpha$-parametrization used. If $\alpha$ is not a constant, but dependent on disk temperature for example, the dependence of the thermal time scale on disk properties will become apparent again.

If, as seems likely from observations, $\alpha$ is generally $<1$, we have in thin disks the ordering of time scales:
\beq t_{\rm v}\gg t_{\rm t}>t_{\rm d}.\eeq

\section{Comparison with CV observations}
The number of meaningful quantitative tests between the theory of disks and
observations is somewhat limited since in the absence of a theory for $\nu$,
it is a bit meagre on predictive power. The most detailed information perhaps
comes from modeling of CV outbursts. 

Two simple tests are possible (nearly) independently of $\nu$. These are the
prediction that the disk is geometrically quite thin (eq.~\ref{hr}) and
the prediction that the surface temperature $T_{\rm s}\sim r^{-3/4}$ in a steady
disk. The latter can be tested in a subclass of the CV's that do not show
outbursts, the nova-like systems, which are believed to be approximately
steady accreters. If the system is also eclipsing, eclipse mapping techniques
can be used to derive the brightness distribution with $r$ in the disk (Horne,
1985, 1993). If this is done in a number of colors so that bolometric
corrections can be made, the results (e.g. Rutten et al. 1992) show in
general a {\it fair} agreement with the $r^{-3/4}$ prediction. Two deviations
occur: i) a few systems show significantly flatter distributions than
predicted, and ii) most systems show a `hump' near the outer edge of the
disk. The latter deviation is easily explained, since we have not taken into
account that the impact of the stream heats the outer edge of the disk.
Though not important for the total light from the disk, it is an important
local contribution near the edge. 

Eclipse mapping of Dwarf Novae in quiescence gives a quite different picture. Here, the inferred surface temperature profile is often nearly flat (e.g. Wood et al. 1989a, 1992). This is understandable however since in quiescence the mass flux depends strongly on $r$. In the inner parts of the
disk it is small, near the outer edge it is close to its average value. With
eq.~(\ref{st}), this yields a flatter $T_{\rm s}(r)$. The lack of light from the
inner disk is compensated during the outburst, when the accretion rate in the
inner disk is higher than average (see Mineshige and Wood 1989 for a more
detailed comparison). The effect is also seen  in the 2-dimensional
hydrodynamic simulations of accretion in a binary by R\'o\.zyczka and Spruit
(1993). These simulations show an outburst during which the accretion
in the inner disk is enhanced, between two episodes in which mass accumulates
in the outer disk.

\section{Comparison with LMXB observations: irradiated disks}
In low mass X-ray binaries a complication arises because of the much higher
luminosity of the accreting object. Since a neutron star is roughly 1000 times
smaller than a white dwarf, it produces 1000 times more luminosity for a given
accretion rate.

Irradiation of the disk by the central source leads to a different surface
temperature than predicted by (\ref{st}). The central source (star plus inner
disk) radiates the total accretion luminosity $GM\dot M/R$ (assuming
sub-Eddington accretion, see section 2). If the disk is {\it concave}, it will
intercept some of this luminosity. If the central source is approximated as a
point source the irradiating flux on the disk surface is
\beq F_{\rm irr}= \epsilon {GM\dot M\over 4\pi Rr^2}, \eeq
where $\epsilon$ is the angle between the disk surface and the direction from a point on the disk surface to the central source:
\beq \epsilon={\mathrm d}H/{\mathrm d}r-H/r.\label{fir}\eeq
The disk is concave if $\epsilon$ is positive. We have
\bd 
{F_{\rm irr}\over F}={2\over 3} {\epsilon\over f} {r\over R},
\ed
where $F$ is the flux generated internally in the disk, given by
(\ref{st}). On average, the angle $\epsilon$ is of the order of the aspect
ratio $\delta=H/r$. With $f\approx 1$, and our fiducial value $\delta\approx
5\,10^{-3}$, we find that irradiation in LMXB dominates for $r > 10^{9}$cm. This is compatible with observations (for reviews see van Paradijs and McClintock 1993), which show that the optical and UV are dominated by reprocessed radiation. 

When irradiation by an external source is included
in the thin disk model, the surface boundary condition of the radiative transfer problem, equation (\ref{sb}) becomes
\beq 
\sigma_{\rm r} T_{\rm s}^4=F + (1-a)F_{\rm irr}, \label{sti}
\eeq
where $a$ is the X-ray albedo of the surface, i.e. $1-a$ is the fraction of the
incident flux that is absorbed in the {\it optically thick} layers of the disk
(photons absorbed higher up only serve to heat up the corona of the disk). The surface temperature $T_{\rm s}$ increases in order to compensate for the additional incident heat flux. The magnitude of the incident flux is sensitive to
the assumed disk shape $H(r)$, as well as on the assumed shape (plane or spherical, for example) of the central X-ray emitting region. The disk thickness depends on temperature, and thereby also on the irradiation. It turns out, however, that this dependence on the irradiating flux is small, if the disk is optically thick, and the energy transport is by radiation (Lyutyi and Sunyaev 1976). To see this, integrate (\ref{edd}) with the modified boundary condition (\ref{sti}). This yields
\beq
\sigma_{\rm r} T^4={3\over 4}F(\tau+{2\over 3})+{(1-a)F_{\rm irr}\over F}.
\eeq
The irradiation adds an additive constant to $T^4(z)$. At the midplane, this constant has much less effect than at the surface. For the midplane temperature and the disk thickness to be affected significantly, it is necessary that
\beq F_{\rm irr}/F \gapprox \tau.\eeq
The reason for this weak dependence of the midplane conditions on irradiation is the same as in radiative envelopes of stars, which are also insensitive to the surface boundary condition. The situation is very different for convective disks. As in fully convective stars, the adiabatic stratification then causes the conditions at the midplane to depend much more directly on the surface temparture. The outer parts of the disks in LMXB with wide orbits may be convective, and their thickness affected by irradiation.

In the reprocessing region of the disks of LMXB, the conditions are such that $F<< F_{\rm irr} \approx \tau F$, hence we must use eq.~(\ref{hr}) for $H$. This yields $\epsilon= (21/20)H/r\approx 5\,10^{-3}$, and $T_{\rm s}\sim r^{0.5}$, and we still expect the disk to remain thin. 

From the paucity of sources in which the central source is eclipsed
by the companion one deduces that the companion is barely or not at all 
visible from the inner disk, presumably because the outer parts of the disk
are much thicker than expected from the above arguments. This is consistent
with the observation that the characteristic modulation of the optical
light curve due to irradiation of the secondary's surface by the X-rays is
not very strong in LMXB (with the exception of Her X-1, which has a large
companion). The place of the eclipsing systems is taken by the so-called
`Accretion Disk Corona' (ADC) systems, where shallow eclipses of a rather
extended X-ray source are seen instead of the expected sharp eclipses of the inner disk (for reviews of the observations, see Lewin et al. 1995). The conclusion is that there is an extended X-ray scattering `corona' above the disk. It scatters a few per cent of the X-ray luminosity.

What causes this corona and the large inferred thickness of the disk ? The thickness expected from disk theory is a rather stable
small number. To `suspend' matter at the inferred height of the disk forces are
needed that are much larger than the pressure forces available in an
optically thick disk. A thermally driven wind, produced by X-ray heating of the
disk surface, has been invoked (Begelman et al. 1983,  Schandl and
Meyer 1994). For other explanations, see van Paradijs and McClintock (1995).
Perhaps a magnetically driven wind from the disk, such as seen in protostellar
objects (e.g. K\"onigl and Ruden 1993) can explain both the shielding of the
companion and the scattering. Such a model would resemble  magnetically driven wind models for the broad-line region in AGN (e.g. Emmering et al., 1992, K\"onigl and Kartje 1994). A promising possibility is that the reprocessing region at the disk edge consists of matter `kicked up' at the impact of the mass transfering stream (Meyer-Hofmeister et al.1997, Armitage and Livio 1998, Spruit et al. 1998). This produces qualitatively the right dependence of X-ray absorption on orbital phase in ADC sources, and the light curves of the so-called supersoft sources.

\subsection{Transients}
Soft X-ray transients (also called X-ray Novae) are believed to be binaries
similar to the other LMXB, but somehow the accretion is episodic, with very
large outbursts recurring on time scales of decades (sometimes years). There
are many black hole candidates among these transients (see Lewin et al. 1995 for a review). As with the Dwarf Novae, the time dependence of
the accretion in transients can in principle be exploited to derive
information on the disk viscosity, assuming that the outburst is caused by an
instability in the disk. The closest relatives of soft transients among the
White Dwarf plus main sequence star systems are probably the WZ Sge stars
(van Paradijs and Verbunt 1984, Kuulkers et al. 1996), which show 
(in the optical) similar outbursts with similar recurrence times (cf. Warner 
1987, O'Donoghue et al. 1991). Like the soft transients, they have low mass 
ratios ($q<0.1$). For a given angular momentum loss, systems with low mass 
ratios have low mass transfer rates, so the speculation is that the peculiar 
behavior of these systems is somehow connected with a low mean accretion 
rate. 

\subsection{Disk Instability}
\label{inst}
The most developed model for outbursts is the disk instability model of Osaki
(1974), H\= oshi (1979), Smak (1971, 1984), Meyer and Meyer-Hofmeister (1981), see also King (1995), Osaki (1993). In
this model the instability that gives rise to cyclic accretion is due to a
temperature dependence of the viscous stress. In any local process that causes
an effective viscosity, the resulting $\alpha$- parameter will be a function of
the main dimensionless parameter of the disk, the aspect ratio $H/r$. If this
is a sufficiently rapidly increasing function, such that $\alpha$ is large in
hot disks and low in cool disks, an instability results by the following
mechanism. Suppose we start the disk in a stationary state at the mean
accretion rate. If this state is perturbed by a small temperature increase,
$\alpha$ goes up, and by the increased viscous stress the mass flux $\dot M$
increases. By (\ref{st}) this increases the disk temperature further,
resulting in a runaway to a hot state. Since $\dot M$ is larger than the
average, the disk empties partly, reducing the surface density and the central
temperature (eq. \ref{ts}). A cooling front then transforms the disk to a cool
state with an accretion rate below the mean. The disk in this model switches
back and forth between hot and cool states. By adjusting $\alpha$ in the hot
and cool states, or by adjusting the functional dependence of $\alpha$ on
$H/r$, outbursts are obtained that agree reasonably with the observations of
soft transients (Lin and Taam 1984, Mineshige and Wheeler, 1989). A rather
strong increase of $\alpha$ with $H/r$ is needed to get the observed long
recurrence times. 

Another possible mechanism for instability has been found in 2-D numerical
simulations of accretion disks (Blaes and Hawley 1988, R\'o\.zyczka and
Spruit 1993). The outer edge of a disk is found, in these simulations, to
become dynamically unstable to a oscillation which grows into a strong
excentric perturbation (a crescent shaped density enhancement which rotates at
the local orbital period). Shock waves generated by this perturbation spread
mass over most of the Roche lobe, at the same time the accretion rate onto the
central object is strongly enhanced. This process is different from the
Smak-Osaki-H\=oshi mechanism, since it requires 2 dimensions, and does not
depend on the viscosity (instead, the internal dynamics in this instability
{\it generates} the effective viscosity that causes a burst of accretion).

\subsection{Other Instabilities}
Instability to heating/cooling of the disk can be the due to several effects.
The cooling rate of the disk, if it depends on temperature in an appropriate
way, can cause a thermal instability like that in the interstellar medium.
Other instabilities may result from the dependence of viscosity on conditions
in the disk. For a general treatment see Piran (1978), for a shorter
discussion see Treves et al., 1988.

\section{Sources of Viscosity}
\label{source}
The high Reynolds number of the flow in accretion disks (of the order
$10^{11}$ in the outer parts of a CV disk) would, to most fluid dynamicists,
seem an amply sufficient condition for the occurrence of hydrodynamic
turbulence. A theoretical argument against such turbulence often used in
astrophysics (Kippenhahn and Thomas 1981, Pringle 1981) is that in cool disks
the gas moves almost on Kepler orbits, which are quite stable (except for the
orbits that get close to the companion or near a black hole). This stability is related to the known
stabilizing effect that rotation has on hydrodynamical turbulence (Bradshaw
1969, for a discussion see Tritton 1992). Kippenhahn and Thomas also
point out that the one laboratory experiment that comes close to the situation
in accretion disks, namely the rotating Couette flow, does not become unstable
for parameters like in disks (for the rather limited range in Reynolds numbers
available). A (not very strong) observational argument is that hydrodynamical
turbulence as described above would produce an $\alpha$ that does not depend
on the nature of the disk, so that all objects should have the same value.
This is unlikely to be the case. From the modeling of CV outbursts one knows,
for example, that $\alpha$ probably increases with temperature (more
accurately, with $H/r$, see previous section). Also, there are indications
from the inferred life times and sizes of protostellar disks (Strom et al.
1993) that $\alpha$ may be rather small there, $\sim 10^{-3}$, whereas in
outbursts of CV's one infers values of the order $0.1-1$. 

The indeterminate status of the hydrodynamic turbulence issue is an annoying problem in disk theory. Direct 3-D numerical simulation of the hydrodynamics in accretion disks is possible, and so far has not shown the expected turbulence. In fact, Balbus and Hawley (1996), and Hawley et al (1999) argue, on the basis of such simulations and a physical argument, that disks are actually quite stable against hydrodynamic turbulence, as long as the specific angular momentum increases outward. [Such heresy would not pass a referee in a fluid mechanics journal.] If it is true that disks are stable to hydrodynamic turbulence it will be an uphill struggle to convince the fluid mechanics community, since it can always be argued that one should go to even higher Reynolds numbers to see the expected turbulence in the simulations or experiments.

The astrophysical approach has been to circumvent the problem by finding plausible alternative mechanisms that might work just as well. 
Among the processes that have been proposed repeatedly as sources of viscosity
is convection due to a vertical entropy gradient (e.g. Kley et
al. 1993), which may have some limited effect in convective parts of disks.
Another class are {\it waves} of various kinds. Their effect can be global,
that is, not reducible to a local viscous term because by traveling across
the disk they can communicate torques over large distances. For example, waves
set up at the outer edge of the disk by tidal forces can travel inward and by
dissipating there can effectively transport angular momentum {\it outward}
(e.g. Narayan et al. 1987,  Spruit et al. 1987). A nonlinear version of this
idea are selfsimilar spiral shocks, observed in numerical simulations (Sawada
et al. 1987) and studied analytically (Spruit 1987). Such shocks can
produce accretion at an effective $\alpha$ of $0.01$ in hot disks, but are
probably not very effective in disks as cool as those in CV's and XRB. A
second non-local mechanism is provided by a magnetically accelerated {\it
wind} originating from the disk surface (Blandford 1976, Bisnovatyi-Kogan and
Ruzmaikin 1976, Lovelace 1976, Blandford and Payne 1982, for reviews see Blandford 1989, Blandford and Rees 1992, for an introduction see Spruit 1996). In principle,  such winds can take care of {\it all} the angular momentum loss needed to make accretion possible in the absence of a viscosity (Blandford 1976, K\"onigl 1989). The attraction of this idea is that
magnetic winds are a strong contender for explaining the strong outflows and
jets seen in protostellar objects and AGN. It is not yet clear however if,
even in these objects, the wind is actually the main source of angular
momentum loss. 

In sufficiently cool or massive disks, selfgravitating
instabilities of the disk matter can produce internal friction. Pacz\'ynski
(1978) has proposed that the resulting heating would limit the instability
and keep the disk in a well defined moderately unstable state. The angular
momentum transport in such a disk has been modeled by several authors (e.g.
Ostriker et al. 1999). Disks in XRB are too hot for selfgravity to play a role.

\subsection{magnetic viscosity}
Magnetic forces can be very effective at transporting angular momentum. If it
can be shown that the shear flow in the disk produces some kind of small
scale fast dynamo process, that is, some form of magnetic turbulence, an effective $\alpha\sim O(1)$ expected (Shakura and Sunyaev 1973, Eardley and Lightman 1975, Pudritz 1981, Meyer and Meyer-Hofmeister 1982). Numerical simulations of initially weak magnetic fields in accretion disks have now shown that this does indeed happen in sufficiently ionized disks (Hawley et al. 1995,  Brandenburg et al. 1995, Armitage 1998). These show a small scale magnetic field with azimuthal component dominating (due to stretching by differential rotation). The effective $\alpha$'s are of the order 0.05.  The angular momentum transport is due to magnetic stresses. The fluid motions induced by the magnetic forces contribute only little to the angular momentum transport. In a perfectly conducting plasma this turbulence can develop from an arbitrarily small initial field through magnetic shear instability (also called magetorotational instability, Velikhov 1959, Chandrasekhar 1961, Balbus and Hawley  1991, 1992). The significance of this instability is that it shows that at large conductivity accretion disks must be magnetic. The actual form of the highly time dependent small scale magnetic field which develops can only be found from numerical simulations. 

\subsection{viscosity in radiatively supported disks}
A disk in which the radiation pressure $P_{\rm r}$ dominates must be optically thick (otherwise the radiation would escape).  The radiation pressure then adds to the total pressure is larger than it would be, for a given temperature, if only the  gas pressure were effective. If the viscosity is then parametrized by (\ref{alf}), it turns out (Lightman and Eardley, 1974) that the disk is locally unstable. An increase in temperature increases the radiation pressure, which increases the viscous dissipation and the temperature, leading to a runaway. This has raised the question whether the radiation pressure should be included in the sound speed that enters expression (\ref{alf}). If it is left out, a lower viscosity results, and there is no thermal-viscous runaway. Without knowledge of the process causing the effective viscous stress, this question can not be answered.  Sakimoto and Coroniti (1989) have shown, however, that if the stress is due to some form of magnetic turbulence, it most likely scales with the gas pressure alone, rather than the total pressure. Now that it seems likely, from the numerical simulations, that the stress is indeed magnetic, there is reason to believe that in the radiation pressure-dominated case the effective viscosity will scale as $\nu\sim \alpha P_{\rm g}/(\rho\Omega)$ (this case has not been studied with simulations yet). Nayakshin and Rappaport (1999) show that, depending on how the viscosity scales in the intermediate regime $P_{\rm g}\approx P_{\rm rad}$, interesting cyclic behavior can occur akin to the `S-curve' instability in CV disks (section \ref{inst}).

\section{Beyond thin disks}
Ultimately, much of the progress in developing useful models of accretion
disks will depend on detailed numerical simulations in 2 or 3 dimensions. In
the disks one is interested in, there is usually a large range in length
scales (in LMXB disks, from less than the $10$km neutron star
radius to the more than $10^{5}$km orbital scale). Correspondingly, there is
a large range in time scales that have to be followed. This not technically
possible at present and in the foreseeable future. In numerical simulations
one is therefore limited to studying in an approximate way aspects that are
either local or of limited dynamic range in $r,t$ (for examples, see Hawley
1991, R\'o\.zyczka and Spruit 1993, Armitage 1998). For this reason, there is still a need
for approaches that relax the strict thin disk framework somewhat without
resorting to full simulations. Due to the thin disk assumptions, the pressure gradient does not contribute to the support in the radial direction and the transport of heat in the radial direction is negligible. Some of the physics of thick disks can be included in a fairly consistent way in the  `slim disk' approximation (Abramowicz et al., 1988). The so-called Advection Dominated Accretion Flows (ADAFs) are related to this approach (for a review see Yi 1998, for an introduction Spruit, elsewhere in this volume).

\subsection{Boundary layers}
\label{bl}
In order to accrete onto a star rotating at the rate $\Omega_*$, the disk matter must dissipate an amount of energy given by
\beq 
{GM\dot M\over 2R}\left[1-\Omega_*/\Omega_k(R)\right]^2. \label{wb}
\eeq
The factor in brackets measures the kinetic energy of the matter at the inner
edge of the disk ($r=R$), in the frame of the stellar surface. Due to this
dissipation the disk inflates into a `belt' at the equator of the star, of
thickness $H$ and radial extent of the same order. Equating the radiation
emitted from the surface of this belt to (\ref{wb}) one gets for the surface temperature $T_{\rm sb}$ of the belt, assuming optically thick conditions and a slowly rotating star ($\Omega_*/\Omega_k\ll1$):
\beq 
{GM\dot M\over 8\pi R^2 H}=\sigma_{\rm r} T_{\rm sb}^4
\eeq
To find the temperature inside the belt and its thickness, use eq.~(\ref{t4}).
The value of the surface temperature is higher, by a factor of the order
$(R/H)^{1/4}$, than the simplest thin disk estimate (\ref{st}, ignoring the
$(r/r_i)^{1/2}$ factor). In practice, this works out to a factor of a few.
The surface of the belt is therefore not very hot. The situation is quite
different if the boundary layer is not optically thick (Pringle and Savonije
1979). It then heats up to much higher temperatures. Analytical methods to
obtain the boundary layer structure have been used by Regev and Hougerat
(1988), numerical solutions of the slim disk type by Narayan and Popham (1993), Popham (1997), 2-D numerical simulations by Kley (1991). These considerations are primarily relevant for CV disks; in accreting neutron stars, the dominant effects of radiation pressure have to be included. More analytic progress on the structure of the boundary layer between a disk and a neutron star and the way in which it spreads over the surface of the star is reported by Inogamov and Sunyaev (1999, see also elsewhere in this volume).

\end{document}